# Social Network Extraction Unsupervised

**Mahyuddin K. M. Nasution[*a], RahmadSyah[b]**

[a]Information Technology Study Program, FakultasIlmuKomputer dan TeknologiInformasi, Universitas Sumatera Utara, Medan 20155, Indonesia
[b]Informatics Department, Universitas Medan Area, Medan, Indonesia
([*]Corresponding author's e-mail: mahyuddin@usu.ac.id)

**Abstract:** In the era of information technology, the two developing sides are data science and artificial intelligence. In terms of scientific data, one of the tasks is the extraction of social networks from information sources that have the nature of big data. Meanwhile, in terms of artificial intelligence, the presence of contradictory methods has an impact on knowledge. This article describes an unsupervised as a stream of methods for extracting social networks from information sources. There are a variety of possible approaches and strategies to superficial methods as a starting concept. Each method has its advantages, but in general, it contributes to the integration of each other, namely simplifying, enriching, and emphasizing the results.
___

## 1. Introduction

By follows the definition of data science [1,2], which confirms that there is a task of disclosing knowledge of big data [3]. Thus, *social network extraction* (*SNE*) from information sources is data modeling that transforms it into knowledge by revealing the existence of social actors and the possible relationships between them [4]. Formally, it states that there is $\gamma: A \times A \to R$ where $A$ is a collection of social actors, and $R$ is a collection of relationships. $\gamma = <\gamma_1, \gamma_2>$ for $\gamma_1(1:1): A \to V$ and $\gamma_2: R \to E$ reveals that $\gamma$: *SNE* $(A,R) \to G(V,E)$ denotes the extraction of social networks with $v_i = \gamma_1(a_i)$ where $v_i$ in $V$ and $a_i$ in $A$, $i = 1,\ldots,N$ and where $e_j = \gamma(r_k(a,b)) = \gamma_2(z_a \cap z_a)$, $e_j$ in $E$, $j=1,\ldots,M$, $r_k$ in $R$ for each $a,b$ in $A$, $k = 1,\ldots,K$ where $z_a \subset Z$, $z_b \subset Z$, and $z_a \cap z_a \subset Z$ and $Z$ are semantic sets of interpretation clues [5,6].

One of the streams of the SNE method is the unsupervised method [7,8,9], which is a superficial way of interpretation that produces knowledge about social structures based on information sources [10]. That is a method that generates interpretations that rely on search engines and the meaning of query results [11]. However, this meaning can be enriched through different strategies, of course, resulting in different resultants. Therefore, it is possible to reveal a variety of social networks that then form different communities from different social structures. This article aims to describe the possible changes in the SNE strategy in an unsupervised method stream.

## 2. Materials and methods

To get information from big data (or information source) that has easy access involves a search engine [12], where big data $\Omega$ is a platform consisting of a collection of data with various characteristics [13]. Thus, for the query $q$, the parts of big data, namely $\omega$ in $\Omega$, reveal the relevance between $q$ and $\omega$. The logical implication $\omega \Rightarrow q$ means that for $q = t_a$, or a query containing the search term $t_a$, where $t_a$ represents the name of the social actor [14]. The meaning of $\omega \Rightarrow t_a$ gives the cumulative as $\Sigma t_a$ against $\Omega_a$ if $t_a$ is true at $\omega$ in $\Omega$ where $|\Omega|$ is the cardinality of big data whereas $|\Omega_a|$ as a hit count. $\Omega_a$ is also called the results of clustering of information space based on the information in the query $q = t_a$ where $\Omega_a \subset \Omega$ and produces not only the hit count but also the snippets, which is a summaries of the information around the query [15,16].

Specifically, on the Internet, the implementation as submission of a query into a search engine produces a web snippet (snippet) that contains at least three elements, namely the URL (uniform resources locator) address of the web page $\omega$, the web page header (title), and the summary of information [17]. The title of web is an important piece of information from a summery point of view, while the summary is part of the web page body. Snippet generally contains ±50 words around the query $q$. Thus, number of words in a snippet is $|s| = \pm 50$ words [18].

Each query $q$ against search engine such as Yahoo, Google, Bing, and others, generally generates a set of snippets according to the size of the snippet value or the hit count. In other words, for $|\Omega_a| > 0$, the number of snippets around $t_a$ is $|\Omega_a|$. It means that there is $|\Omega_a|$ URL address, $|\Omega_a|$ titles of web page, and $|\Omega_a|$ summery of web pages [19]. In other words, for each $t_a$ and $|\Omega_a| = l$ there is a collection of snippets $L_a = \{S_t|t=1,\ldots, l, t_a \text{ in } S_t\}$. Suppose $w$ is a word, $w$ in $\omega$ and for $|\Omega_a| > 0$ then $L_a$ is a bag of words (BoW) [20]. Technically every word on a web page has a value, that is $p(w) = |w|/|s|$, which is the number of the same words $|w|$ compared to the total number of words on the snippet $|s|$. Suppose $|w|_t$ is the number of same words in the $t$-th snippet, the value of each word $w$ in BoW [21,22,23], that is $p(w_h)$, i.e.,

$$p(w_h) = (\sum_{t=1,\ldots,l}|w|_t/|s|_t)/t \qquad (1)$$

$h = 1,\ldots,H$, $H$ is number of words uniquely in BoW. The normalized value $pr(w)$ is calculated as

$$pr(w) = p(w_h)/\text{sqrt}(\sum_{u=1,\ldots,U}p(w_h)^2) \qquad (2)$$

Similarly, for a query $q = t_a,t_b$, or a query that forms a cluster of information as $\Omega_a \cap \Omega_b$, if hit count $|\Omega_a \cap \Omega_b| > 0$ then there are one or more snippets about the contents of $q$. The BoW of $\Omega_a \cap \Omega_b$ means that there are $p(w_u)$ and $pr(w)$ for $w$ in $\Omega_a \cap \Omega_b$ for $a,b$ in $A$. Therefore, the occurrence $\Omega_a$, occurrence $\Omega_b$, co-occurrence $\Omega_a \cap \Omega_b$, snippet $S$, word $w$, and others together are clues of $Z$ [25].

The extracting social networks [26] from conversational speech transcripts [27], Chinese news [28], literary fiction [29], mobile phone PSP messages [30], transaction logs [31], language statistics [32], e-government [33]. the Web [8,10], by using unsupervised methods [34,35]. The extracting which is specifically automatic [36], namely to reveal the existence of social actors and then reveal the presence of the relations between them [37,38]. Usually, the relationship between two social actors is modeled by grouping actors through similarity measures [39]. The measurement includes the formula for the Jaccard coefficient, i.e.

$$J_c = |\Omega_a \cap \Omega_b|/(|\Omega_a|+|\Omega_b|-|\Omega_a \cap \Omega_b|) \qquad (3)$$

and other similarity measures such as cosine, mutual information, overlap coefficient, dice coefficient, etc., which are generally expressed as $sr = sim(|\Omega_a|,|\Omega_b|,|\Omega_a \cap \Omega_b|)$ [40,41]. So, to build the strength relation ($sr$) between a pair of social actors is as expressed in the algorithm in the following procedure [42]:

**Algoritma 1**:

function BSM($a,b$) {

    $|\Omega_a \cap \Omega_b| \leftarrow \text{query}(t_a,t_b)$

    if $|\Omega_a \cap \Omega_b|>0$ then

        $|\Omega_a| > 0 \leftarrow \text{query}(t_a)$

        $|\Omega_b| > 0 \leftarrow \text{query}(t_b)$

        if $|\Omega_a|>=|\Omega_a \cap \Omega_b|$ and $|\Omega_a|>=|\Omega_a \cap \Omega_b|$ then

            $sr = sim(|\Omega_a|,|\Omega_b|,|\Omega_a \cap \Omega_b|)$

    else $sr = 0$

}

return $sr$

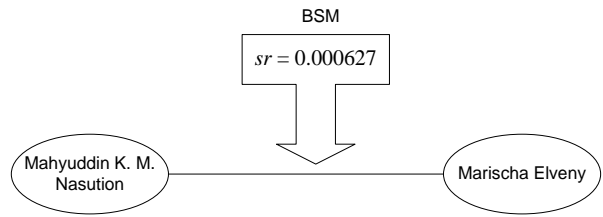

**Figure 1**. Strength relation between social actors based on BSM.

The naming of Algorithm 1 is the *basic superficial method* (BSM), which reveals the strength of the relation between two social actors. For example, for a social actor $t_a$ = Mahyuddin K. M. Nasution based on a Google search engine, the query returns $|\Omega_a|$ = 121,000 hits, while for a social actor $t_b$ = MarischaElveny the query returns $|\Omega_b|$ = 2,130,000 hits. In semantics, the search for each social actor's name is declared as an occurrence. Every pair of occurrences to produce a relationship between social actors semantically involves a co-occurrence directly due to the existence of $|\Omega_a \cap \Omega_b|$ = 1,410 hits as a result of the query $q = t_a,t_b$ [43,44]. Then $sr$ becomes 0.000627, or as shown in Figure 1, a graph of a social network [45].

## 3. Results and discussion

The automatic implementation of the BSM on a basis can involve the Python programming language, for example, to access information using the yahoo search engine [46], for example:

page = urlopen("http://search.yahoo.com/search;...nama1,nama2...").read()

where "..." is the completeness of the query that directs nama1 and nama2 as the information sought. Two variables nama1 and nama2 contain the names of social actors based on the query concept $q=t_a,t_b$, and by involving the hit count metadata search commands from the return results of the search engine on the page has a co-occurrence. Usually, a page apart from containing the hit count also consists of one web page containing 10 snippets maximum even though $|\Omega_a \cap \Omega_b|$ greater than 10 hits [47]. The page consists of a URL address, header, and summary of the web page mixed with tags. To produce plaintext is by reducing tags of the web the Python programming language has equipment with special natural language processing functions that make it easy to generate BoW from a snippet page. It is to enrich the social structure [48].

Strategy change in BSM is dealing with query content submitted to search engine. Using well-defined social actor names in quotation marks gives a different hi count. For example, $q$ = "Mahyuddin K. M. Nasution","MarischaElveny" involving the Google search engine produces $|\Omega_{"a"} \cap \Omega_{"b"}|$ = 774 where $|\Omega_{"a"}|$ = 6,740 and $|\Omega_{"b"}|$ = 2,470. By involving Eq. (3) yields $sr$ = 0.09175. This change in strategy affects the method of extracting social networks from information sources. Call it as *the pattern superficial method* (PSM). It aims to emphasize the strength relation between social actors [18].

One of the contents of the snippet is the URL address of a web page where the query content is located. The canonical form of URL consists of components in $U$ = {$s,d,p,q$} = {*scheme,authority,path, query*}, which form the string $s://d_m.\cdots.d_2.d_1/p_1/p_2/\ldots/p_{n-2}/x$, $x = p_{n-1}$ or $x = p_{n-1}?q$ [18]. Thus the URL has n layers and separating each part by a slash. Each occurrence indirectly presents maximal ten URL addresses on one page for $|\Omega_a| > 0$ or $|\Omega_{"a"}| > 0$, for example. The similarity between URL addresses of two occurrences can form a relationship between two social actors. Snippets of $q = t_a$ have $|\Omega_a|$ URL addresses, as well as snippets of $q = t_b$ produce $|\Omega_b|$ URL address. Two sets of URL addresses, it is possible that there is the same URL address between the two, similar URL addresses between the two, or different URL addresses. The exact same URL address forms a strong semantic relationship, in which the co-occurrence of the names of social actors is on the same page. Similar URL addresses reveal that the initial layers of the URL are the same, while the ending is different, and it semantically reveals the co-occurrence of social actor names on the same page of a document but between the two social actor names there is a three-point barrier "…", which implies a different meaning of the relationship [38]. For example, if an author is citing scientific work from another author, both authors are usually limited by "…".By parsing the URL layers from dissimilar to the same layer, it is possible to construct a similarity and involve measurements from Eq. (3). Or to construct a new similarity adapting the URL address, for example, $sim(a,b) = 2|ab|/(|a|+|b|)$, $sim(a,b)$ in [0,1] [49], where $|ab|$ is the cardinality of $|a \cap b|$, while $|a|$ and $|b|$ respectively the cardinality of the vectors of $t_a$ and $t_b$. It is an approach that involves changing this strategy. Let's call it as the *underlying superficial method* (USM) [10], have other variations. The first variation when the query involves a pattern of names, where the URL address in the snippet is in $\Omega_{"a"}$, for example. The initial variation or USM becomes the *basic underlying superficial method* (bUSM), while another variation involving the pattern becomes the *pattern underlying superficial method* (pUSM). In contrast, co-occurrence will result in a list of URLs that are the same between the two social actors. Thus, the cardinality of co-occurrence is proportional to $|\Omega_a \cap \Omega_b|$, which means that there is $|\Omega_a \cap \Omega_b|$ URL address on-page. Comparative formula between $|\Omega_a \cap \Omega_b|$ and $n$ layers of URL addresses, namely $n/(|\Omega_a \cap \Omega_b|)$ apply $|\Omega_a \cap \Omega_b| > n$, $|\Omega_a \cap \Omega_b| = n$, or $|\Omega_a \cap \Omega_b| < n$. To get the strength relation requires normalization so that $sr$ in [0,1]. This strategy is also known as cbUSM while another variation is cpUSM, by changing the charge from $\Omega_a \cap \Omega_b$ to $\Omega_{"a"} \cap \Omega_{"b"}$.

For example, one of the URL addresses in the snippet list for $t_a$ is

https://publons.com/researcher/2908750/mahyuddin-k-m-nasution/

while in the list of snippets for $t_b$ is

https://publons.com/researcher/1730428/marischa-elveny/

each has four layers, both URL addresses that are not exactly the same but have similarities. In addition to the two initial layers associated with the Publon site and the naming of the researcher community, the next two layers each provide researched and researcher names. Each $|a|$ and $|b|$ while the value is four, whereas $|ab|$ has a value of 2, so $sim(a,b) = 2|ab|/(|a|+|b|) = 2(2)/(4+4) = 4/8 = 0.5$ [10,49]. Accumulatively for all URLs in the snippet of either $t_a$ or $t_b$, it will produce $|a|$ and $|b|$, in the same way, the accumulative value for $|ab|$. If the measurement involves vector values of |*a*|, |*b*|, and |*ab*|, then for each of the approaches from bUSM, pUSM, cbUSM, and cpUSM have implemented different strategies and produced four measures of *sr*.

Based on both titles and summary of web pages in the snippet provide a set of words. The sets come from the occurrence of each author or the co-occurrence of two authors, where each word in BoW has a value according to Eq. (2) [5]. A set of words according to each social actor based on occurrence produces a vector as |*a*| or |*b*| respectively of $\Omega_a$ and $\Omega_b$, while the same set of words based on the occurrence of a pair of social actors produces a vector as |*ab*| of $\Omega_a$ and $\Omega_b$, or $|ab|_{2o}$. In contrast, the co-occurrence $\Omega_a \cap \Omega_b$ pair of social actors also produces vectors such as |*ab*|, or $|ab|_c$. $|ab|_{2o}$ is not always the same value as $|ab|_c$. So, there is a variation of the measurement that results in a variation of the strength relation, *sr*, which involves measuring |*a*|, |*b*|, and $|ab|_{2o}$, measurement with |*a*|, |*b*|, and $|ab|_c$, and measurement only involve $|ab|_c$ by constructing the normalization of $|ab|_c$ based on their mean values. When the measurement strategies vary according to the data model, there are various adaptation methods, namely the *occurrence description superficial method* (oDSM), *basic description superficial method* (bDSM), and *co-occurrence description superficial method* (cDSM). By changing the content of the query, i.e., involves a well-defined name, also there are indirectly three different variations, namely the *pattern occurrence description superficial method* (poDSM), the *pattern basic description superficial method* (pbDSM), and the *pattern co-occurrence description superficial method* (pcDSM) [14,18].

**Table 1.** Methods in an unsupervised stream for extracting social networks from the Web.

| | | Generate and Reveal | | | | | | |
|---|---|---|---|---|---|---|---|---|
| *ij* | Relationships | Basic | Underlying | | Descriptive | | | Average |
| | Author - co-Author | BSM | bUSM | cbUSM | bDSM | oDSM | cDSM | $\mu$ |
| 1 | Mahyuddin K. M. Nasution - MarischaElveny | $sr_1$ | $sr_2$ | $sr_3$ | $sr_4$ | $sr_5$ | $sr_6$ | $\mu_1$ |
| 2 | … | $sr_1$ | $sr_2$ | $sr_3$ | $sr_4$ | $sr_5$ | $sr_6$ | $\mu_2$ |
| 3 | … | $sr_1$ | $sr_2$ | $sr_3$ | $sr_4$ | $sr_5$ | $sr_6$ | $\mu_3$ |
| | | Enrichment and Confirmation | | | | | | |
| *ij* | Relationships | Pattern | Underlying | | Descriptive | | | Average |
| | Author - co-Author | PSM | pUSM | cpUSM | pbDSM | poDSM | pcDSM | $\eta$ |
| 1 | "Mahyuddin K. M. Nasution" – "MarischaElveny" | $sr_7$ | $sr_8$ | $sr_9$ | $sr_{10}$ | $sr_{11}$ | $sr_{12}$ | $\eta_1$ |
| 2 | … | $sr_7$ | $sr_8$ | $sr_9$ | $sr_{10}$ | $sr_{11}$ | $sr_{12}$ | $\eta_2$ |
| 3 | … | $sr_7$ | $sr_8$ | $sr_9$ | $sr_{10}$ | $sr_{11}$ | $sr_{12}$ | $\eta_3$ |

Social network extraction methods in the unsupervised stream generally prioritize the means of access to the available information space or big data. Those access tools, for example, search engines, of course, involve queries. Besides, access to the information space is through the available log system and the granting of authorization to the information space or database. Several access tools provide easy entry to specific information spaces through strategies that involve special formulations [5]. To reveal the relationship between social actors besides getting the occurrence and co-occurrence, it also involves a measurement which results in the value being in [0,1]. Usually, it uses similarity but does not rule out using the average from the measurement. The methods always accompany by a way to evaluation approach by disclosing information through surveys and involves measuring recall, precision, or F-measure [44,50].

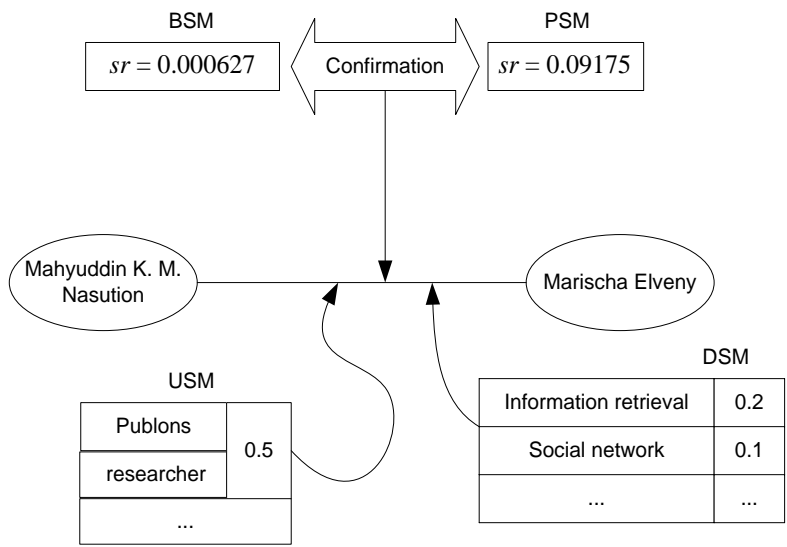

**Figure 2.** Enrichment and confirmation of strength relation.

All the approaches and strategies that make up the method reveal that there is the simplest method, and other methods can enrich by involving additional information that explains the formation of a relationship or community. Then, each relationship has a confirmation. Usually, its importance involves a threshold. Based on Table 1, there is an average formulation,

$$\mu = (\Sigma_{k=1\ldots6} sr_k)/6 \tag{4}$$

Eq. (4) integrates measurement by completing the initial enrichment, meanwhile

$$\eta = (\Sigma_{k=7\ldots12} sr_k)/6 \tag{5}$$

It integrates measurement by asserting measurement and enrichment, Figure 2. So, integrated measurement is $\mu + \eta$ in unsupervised stream.

## 4. Conclusions

Social network extraction involves access tools, search engines, queries, hit counts, similarity measurements, generally recognized as superficial methods. This method by changing the strategy provides enrichment and confirmation of the measurement results. These methods becomes important in the extraction of social networks from information sources. The next task of this research is to reveal the complexity of the social network extraction method

**References**

G Klepac, R Kopal, L Mršíc. Social network metrics integration into fuzzy expert system and Bayesian network for better data science solution performance. Hybrid Intelligence for Social Networks, 25-45. DOI: 10.1007/978-3-319-65139-2_2.

M K M Nasution, O S Sitompul, E B Nababan. Data Science. Journal of Physics: Conference Series, 2020, 1566(1). DOI: 10.1088/1742-6596/1566/1/012034.

V Grossi, B Rapisarda, F Giannotti, D Pedreschi. Data science at SoBigData: the European research infrastructure for social mining and big data analytics. International Journal of Data Science and Analytics, 2018, 6(3), 205-216.

A Farasat, G Gross, R Nagi, A G Nikolaev. Social network extraction and high value individual (HVI) identification within fused intelligence data. Lecture Notes in Computer Science (including subseries Lecture Notes in Artificial Intelligence and Lecture Notes in Bioinformatics), 2015, 9021, 44-54.

M K M Nasution, S A Noah. Extraction of academic social network from online database. 2011 International Conference on Semantic Technology and Information Retrieval, STAIR 2011, 64-69. DOI: 10.1109/STAIR.2011.5995766.

M K M Nasution. Social network mining (SNM): A definition of relation between the resources and SNA. International Journal on Advanced Science, Engineering and Information Technology, 2016, 2(6), 975-981. DOI: 10.18517/ijaseit.6.6.1390.

Y Jin, Y Matsuo, M Ishizuka. Extracting social networks among various entities on the Web. Lecture Notes in Computer Science (including subseries Lecture Notes in Artificial Intelligence and Lecture Notes in Bioinformatics), 2007, 4519 LNCS, 251-266. DOI: 10.1007/978-3-540-72667-8_19,

F Dellutri, L Laura, V Ottaviani, G F Italiano. Extracting social networks from seized smartphones and web data. Proceedings of the 2009 1st IEEE International Workshop on Information Forensics and Security, WIFS 2009, 2009, 101-105. DOI: 10.1109/WIFS.2009.5386473.

N Spirin, K Karahalios. Unsupervised approach to generate informative structured snippets for job search engines. WWW 2013 Companion – Proceedings of the 22nd International Conference on World Wide Web, 2013, 203-204. DOI: 10.1145/2487788.2487891.

M K M Nasution, S A Noah. Superficial method for extracting social network for academics using web snippets. Lecture Note in Computer Science (including subseries Lecture Notes in Artificial Intelligence and Lecture Notes in BioInformatics), 2010, 6401 LNAI, 483-490. DOI: 10.1007/978-3-642-16248-0_68.

S Siersdorfer, P Kemkes, H Ackermann, S Zerr. Who with whom and how? – Extracting large social networks using search engine. International Conference on Information and Knowledge Management, Proceedings, 2015, 1491-1500. DOI: 10.1145/2806416.2806582.

C H Li. Search engine intelligent algorithm for big data. Telecommunication and Radio Engineering (English translation of Elektrosvyaz and Radiotekhnika), 2020, 79(10), 883-890. DOI: 10.1615/TelecomRadEng.v79.i10.50.

I C Drivas, D P Skas, G A Giannakopoulos, D. Kyriaki-Manessi. Big data analytics for search engine optimization. Big Data and Cognitive Computing, 2020, 4(2), 1-22.

M K M Nasution, S A Noah. Social network extraction based on Web. A comparison of superficial methods. Procedia Computer Science, 2017, 124, 86-92. DOI: 10.1016/j.procs.2017.12.133.

T Penin, H Wang, T Tran, Y Yu. Snippet generation for semantic web search engines. Lecture Notes in Computer Science (Including subseries Lecture Notes in Artificial Intelligence and Lecture Notes in BioInformatics), 2008, 5367 LNCS, 493-507. DOI: 10.1007/978-3-540-89704-0_34.

P Ferragina, A Gulli. A personalized search engine based on web-snippet hierarchical clustering. 14th International World Wide Web Conference, WWW2005, 2005, 801-810. DOI: 10.1145/1062745.1062760.

A Strzelecki, P Rutecka. The snippets taxonomy in Web Search Engines. Lecture Notes in Business Information Processing, 2019, 365, 177-188. DOI: 10.1007/978-3-030-31143-8_13.

M K M Nasution. Social network extraction based on Web: 1. Related superficial methods. IOP Conference Series: Materials Science and Engineering, 2018, 300(1). DOI: 10.1088/1757-899X/300/1/012056.

M K M Nasution. Singleton: A role of the search engine to reveal the existence of something in information space. IOP Conference Series: Materials Science and Engineering, 2018, 420(1). DOI: 10.1088/1757-899X/420/1/012137.

K Maruyama, M Igeta, M Terada. Search engine result page with visual context and already rendered snippets. WEBIST 2011 – Proceedings of the 7th International Conference on Web Information Systems and Technologies, 2011, 340-345.

K T Ahmed, H Afzal, M R Mufti, A Mehmood, G S Choi. Deep image sensing and retrieval using suppression, scale spacing and division, interpolation and spatial color coordinates with bag of words for large and complex datasets. IEEE Access, 2020, 8, 90351-90379. DOI: 10.1109/ACCESS.2020.2993721.

R S A Ameer, M Al-Taei. Human action recognition based on bag-of-words. Iraqi Journal of Science, 2020, 61(5), 1202-1214. DOI: 10.24996/ijs.2020.61.5.27.

M Klungpornkun, P Vateekul. Hierarchical text categorization using level based neural networks of word embedding sequences with sharing layer information. Walailak Journal of Science & Technology, 2019, 16(2), 121-131. DOI: https://doi.org/10.14456/vol16iss2pp%25p.

Q Zhou, L Leydesdorff. The normalization of occurrence and co-occurrence matrices in bibliometrics using cosine similarities and ochiaicofficients. Journal of the Association for Information Science and Technology, 2016, 67(11). 2805-2814. DOI: 10.1002/asi.23603.

M K M Nasution. Doubleton: A role of the search engine to reveal the existence of relation in information space. IOP Conference Series: Materials Science and Engineering, 2018, 420(1). DOI: 10.1088/1757-899X/420/1/012138.

M Forestier, J Velcin, D Zighed. Extracting social networks to understand interaction. Proceedings – 2011 International Conference on Advances in Social Networks Analysis and Mining, ASONAM, 2011, 213-219. DOI: 10.1109/ASONAM.2011.64.

H Jing, N Kambhatla, S Roukos. Extracting social networks and biographical facts from conversational speech transcripts. ACL 2007 – Proceedings of the 45th Annual Meeting of the Association for Computational Linguistics, 2007, 1040-1047.

Y Weijie, D Ruwei, C Xia. Extracting social network among various entities from Chinese news stories by content analysis. Proceedings – International Computer Software and Applications Conference, 2008, 929-934. DOI: 10.1109/COMPSAC.2008.6.

D K Elson, N Dames, K R McKeown. Extracting social networks from literary fiction. ACL 2010 – 48th Annual Meeting of the Association for Computational Linguistics, Proceedings of the Conference, 2010, 138-147.

K Xu, Y Chen, K Zou. Extracting social networks from mobile phone PSP messages. Communication in Computer and Information Science, 2012, 267 CCIS, Issue PART 1, 208-214. DOI: 10.1007/978-3-642-29084-8_31.

C Chen, B Xu, Y Xiao, Q Shi, W Wang. Extracting social network from transaction logs. JisuanjiYanjiuyuFazhan/Computer Research and Development, 2015, 52(11), 2508-2516. DOI: 10.7544/issn1000-1239.2015.20148134.

S Hutchinson, M Louwerse. 2018. Extracting social networks from language statistics. Discourse Processes, 2018, 55(7), 607-618. DOI: 10.1080/0163853X.2017.1332446.

R M Alguliyev, R M Aliguliyev, G Y Niftaliyeva. Extracting social networks from e-government by sentiment analysis of user' comments. Electronic Government, 2019, 15(1), 91-106. DOI: 10.1504/EG.2019.096576.

O I Osesina, J P McIntire, P R Havig, E EGeiselman, C Bartley, M E Tudoreanu. Methods for extracting social network data from chatroom logs. Proceedings of SPIE – The International Society for Optical Engineering, 2012, 8389. DOI: 10.1117/12.920019.

M Elfida, M K M Nasution, O S Sitompul. Enhancing to method for extracting social network by the relation existence. IOP Conference Series: Materials Science and Engineering, 2018, 300(1). DOI: 10.1088/1757-899X/300/1/012057.

B Turnbull, S Randhawa. Automatic event and social network extraction from digital evidence sources with ontological mapping. Digital Investigation, 2015, 13, 94-106. DOI: 10.1016/j.diin.2015.04.004.

M K M Nasution. New method for extracting keyword for the social actor. Lecture Notes in Computer Science (including subseries Lecture Notes in Artificial Intelligence and Lecture Notes in Bioinformatics), 2014, 8397 LNAI, Issue PART 1, 83-92. DOI: 10.1007/978-3-319-05476-6_9.

M K M Nasution, O S Sitompul. Enhancing extracting method for aggregating strength relation between social actors. Advances in Intelligent Systems and Computing, 2017, 573, 312-321. DOI: 10.1007/978-3-319-57261-1_31.

B Saxena, V Saxena. Towards establishing the effect of self-similarity on influence maximization in online social networks. Social Network Analysis and Mining, 2020, 10(1). DOI: 10.1007/s13278-020-00654-7.

M K M Nasution. Social network extraction based on Web: 2. Strategies in superficial methods. Journal of Physics: Conference Series, 2018, 1116(2). DOI: 10.1088/1742-6596/1116/2/022029.

M K M Nasution, O S Sitompul, M Elveny, R Syah, R F Rahmat. A similarity for new meanings. 2020 International Conference on Data Science, Artificial Intelligence, and Business Analytics, DATABIA 2020 – Proceedings, 2020, 37-44. DOI: 10.1109/DATABIA50434.2020.9190316.

M K M Nasution, S A M Noah. Comparison of the social network weight measurements. IOP Conference Series: Materials Science and Engineering, 2020, 725(1). DOI: 10.1088/1757-899X/725/1/012094.

M K M Nasution, O S Sitompul, S A Noah. Social network extraction based on Web: 3. The integrated suprficial method. Journal of Physics: Conference Series, 2018, 978(1). DOI: 10.1088/1742-6596/978/1/012033.

M K M Nasution, S A M Noah. Social network extraction based on Web: 4. A framework. Journal of Physics: Conference Series, 2020, 1566(1). DOI: 10.1088/1742-6596/1566/1/012029.

F M Clemente, F M L Martins. Who are the prominent players in the UEFA champions league? An approach based on network analysis. Walailak Journal of Science & Technology, 2017, 14(8), 627-636. https://doi.org/10.14456/vol14iss8pp%25p

M K M Nasution. 2013 Superficial method for extracting academic social network from the Web. Ph.D. Dissertation. UniversitiKebangsaan Malaysia, Selangor, Malaysia.

P Murugesan, K Malathi. Efficient search engine approach for measuring similarity between words: Using page count and snippets. IC-GET 2015 – Proceedings of 2015 Online International Conference on Green Engineering and Technologies, 2015. DOI: 10.1109/GET.2015.7453830.

M Forestier, J Velcin, D Zighed. Extracting social networks enriched by using text. Lecture Notes in Computer Science (including subseries Lecture Notes in Artificial Intelligence and Lecture Notes in Bioinformatics), 2011, 6804 LNAI, 140-145. DOI: 10.1007/978-3-642-21916-0_16.

M K M Nasution, O S Sitompul. S Nasution, H Ambarita. New similarity. IOP Conference Series: Materials Science and Engineering, 2017, 180(1). DOI: 10.1088/1757-899X/180/1/012297.

M K M Nasution, S A Noah. Information retrieval model: A social network extraction perspective. Proceedings – 2002 International Conference on Information Retrieval and Knowledge Management, CAMP'12, 2012, 322-326. DOI: 10.1109/InfRKM.2012.6204999.